\documentclass[11pt]{article}
\usepackage{amsfonts} \usepackage{amssymb} 
\newcommand{\mrm}[1]{\mbox{\rm #1}}
\newcommand{\nn}{\nonumber}
\newcommand{\be}{\begin{equation}}
\newcommand{\bea}{\begin{eqnarray}}
\newcommand{\eea}{\end{eqnarray}}
\newcommand{\ee}{\end{equation}}
\newcommand{\rfn}[1]{(\ref{#1})}
\newcommand{\eq}[1]{Eq.~(\ref{#1})}

\newcommand{\dv}{\partial\hspace{-7pt}\slash}

\newcommand{\dvl}{\stackrel{\hspace{3pt}\leftarrow}{\partial\hspace{-7pt}\slash}}
\newcommand{\Dr}{\stackrel{\hspace{3pt}\rightarrow}{D\hspace{-8pt}\slash}}
\newcommand{\dvr}{\stackrel{\hspace{3pt}\rightarrow}{\partial\hspace{-7pt}\slash}}

\newcommand{\gev}{ {\rm GeV} }

\begin{document}
%
%
\thispagestyle{empty}
\begin{flushright}
{\tt hep-ph/0210271}\\
{FTUAM-02-26}\\
{IFT-UAM/CSIC-02-46} \\
{UCSD/PTH 02-25}
\end{flushright}
\vspace*{1cm}
\begin{center}
{\Large{\bf The Effective Lagrangian for the Seesaw 
Model of Neutrino Mass and Leptogenesis} }\\
\vspace{.5cm}
A. Broncano$^{\rm a,}$\footnote{alicia.broncano@uam.es}, 
M.B. Gavela$^{\rm a,}$\footnote{gavela@delta.ft.uam.es} and 
E. Jenkins$^{\rm b,}$\footnote{ejenkins@ucsd.edu}
 
\vspace*{1cm}
$^{\rm a}$ Dept. de F\'{\i}sica Te\'orica, C-XI, Facultad de Ciencias, 
Univ. Aut\'onoma de Madrid, Cantoblanco, 28049 Madrid, Spain \\
$^{\rm b}$ Dept. of Physics, University of California at San Diego, 
9500 Gilman Drive, La Jolla, CA 92093, USA

\vspace{.3cm}

%
%
\begin{abstract}
\noindent

The effective Lagrangian for the seesaw model is derived including
effects due to $CP$ violation.
Besides the usual dimension-5 operator responsible for light neutrino 
masses, a dimension-6 operator is obtained. For three or less heavy neutrino 
generations, the inclusion of both operators is necessary 
and sufficient for all independent physical parameters of 
the high-energy seesaw Lagrangian to appear in the low-energy effective 
theory, including the $CP$-odd phases
relevant for leptogenesis.  The dimension-6 operator implies exotic low-energy
couplings for light neutrinos, providing a link between the high-energy physics
and low-energy observables.
 
\end{abstract}

\end{center}
%
%

\pagestyle{plain} 
\setcounter{page}{1}
\setcounter{footnote}{0}

%
\section{Introduction}

There is mounting experimental evidence for neutrino masses and mixings
from oscillation experiments~\cite{experiments}.  It is possible that the light
neutrino masses take natural (i.e. non-fine-tuned) values, in contrast to all other
fermion masses except the top quark mass, if the smallness of neutrino masses is
explained by the seesaw mechanism~\cite{seesaw}.  In the minimal seesaw model~\cite{fy}, 
gauge singlet fermions with Majorana masses of order ${\cal M}$  
much larger than 
the electroweak scale couple to the massless weak doublet neutrinos of the Standard Model
(SM) through Yukawa couplings to the Higgs scalar doublet.
Upon spontaneous symmetry breakdown of the electroweak gauge symmetry, these Yukawa
interactions generate a Dirac neutrino mass between the heavy singlet and weak
doublet neutrinos.  The otherwise massless
weakly-interacting neutrinos develop small masses 
$\sim -m_{\rm Dirac}^2/{\cal M}$. 
The conditions for a seesaw mechanism of light neutrino mass to occur, 
namely the existence of
heavy Majorana gauge singlet fermions with Yukawa couplings to the massless weakly-interacting
neutrinos, naturally arise 
in the context of grand unified theories and partially unified theories.
It is striking that present neutrino oscillation
data indicates a seesaw scale 
${\cal M}$ of new physics that is comparable to the scale at which
the Standard Model gauge couplings are converging.

Present solar and atmospheric neutrino oscillation data imply two
distinct mass differences for  light neutrinos. In a seesaw model, this requires 
at least two heavy 
Majorana singlet neutrinos. As soon as two or more Majorana 
neutrinos are present in the seesaw model, an attractive scenario opens up  
for solving the puzzle of the matter-antimatter asymmetry of the universe:
leptogenesis  at the scale ${\cal M}$ and baryogenesis from the lepton 
asymmetry \cite{fy}.

Recent observations of acoustic peaks in the Cosmic Microwave Background (CMB) \cite{CMB} 
have confirmed and refined the 
estimation of the baryon number $B$ 
in the universe resulting from big bang nucleosynthesis. The ratio of 
the baryon to photon density $\eta = (n_B-n_{\bar{B}})/n_\gamma= n_B/n_\gamma$, extracted from
the CMB, is
\be
\eta= (6.0^{+1.1}_{-0.8}) \times 10^{-10} \nn.
\ee
The Sakharov conditions \cite{Sakharov} for 
baryogenesis are
departure from thermal equilibrium, and violation of $B$, 
$CP$ and $C$. 
Electroweak baryogenesis is ruled out in the SM since $CP$-violation in the quark
sector is insufficient by 
many orders of magnitude \cite{Gavela}, and in addition the Higgs mass seems to be
too large \cite{Csikor}. 
In practice, any efficient mechanism for baryogenesis 
beyond the SM acting at high energies
should break both $B$ and $B$-$L$, where $L$ denotes lepton number; otherwise any baryon asymmetry 
produced at high energy can be washed out by non-perturbative
effects, the sphalerons \cite{sphaleron},  
which preserve $B$-$L$,
well before the electroweak transition \cite{Rubakov}.  It is interesting that
a baryon asymmetry points to ($B$-$L$)-breaking interactions, and hence to $L$-violating
neutrino masses.

An excess of lepton density in the early universe
can be generated dynamically in the seesaw model by decay of the
heavy Majorana neutrinos into light 
leptons and the Higgs boson, since lepton number $L$, $CP$ and $C$ 
are violated by the decay. 
The expansion of the universe naturally provides the necessary 
out-of-thermal-equilibrium condition required for leptogenesis.
The SM interactions recycle about half of this lepton 
asymmetry into a baryon asymmetry by 
active sphaleron processes.  The seesaw model of
leptogenesis is extremely elegant and highly economical,  
since it requires only a minimal extension of the SM to include 
sterile neutrinos.

The experimental discovery of $CP$-violation in the lepton sector would be a 
major breakthrough.  Low-energy Majorana $CP$-odd phases, requiring at 
least two active light neutrino families, may contribute
\cite{CP+beta} to the total 
neutrinoless double beta decay amplitude \footnote{For a recent 
pessimistic appraisal of the practical possibility of extracting them 
in the future, see ref.~\cite{glashow}.}. The ``CKM-like'' $CP$-odd phases, 
requiring at least three active light neutrino families, may be detected 
in neutrino oscillations, provided the large mixing angle MSW solution~\cite{MSW} 
to the solar neutrino deficit is confirmed, and the angle $\theta_{13}$ is not orders 
of magnitude below its present limit~\cite{CHOOZ}. 
It is of prime importance to determine
what is the connection, if any, between the $CP$-odd phases of the 
high-energy seesaw Lagrangian and those to be measured in 
on-going or future experiments.

Establishing whether light neutrino masses are the result of the seesaw model requires
finding an experimental signature of the seesaw model beyond the existence
of light neutrino masses.   
In this work we construct an effective Lagrangian of the seesaw theory
which is valid for energies below the seesaw
scale.
We search for the minimal set of higher dimensional operators, compatible 
with the symmetries of 
the problem, which are necessary to take into account the leading effects involving
the parameters of leptogenesis.
The RG running of the effective
Lagrangian to low energies will be the subject of future 
investigation~\cite{more}.

\section{Integrating out the heavy neutrino} 

We consider the minimal extension of the Standard Model with $n$ light generations in which
$n'$ right-handed neutrinos $N_R$ are added to the field content.  
The most general gauge invariant renormalizable Lagrangian is given by 
\be
\label{Lagrangian}
{\cal L} = {\cal L}_{SM} + {\cal L}_{N_R}\, ,
\ee
where $ {\cal L}_{SM}$ is the SM Lagrangian, while 
\be
{\cal L}_{N_R}= i \overline{N_R}\, \dv \ N_R
- \left( \overline{\ell_L} \,{\widetilde\phi} \, {Y_\nu}  \, N_R  
+\overline{N_R} \, {Y_\nu}^\dagger \, {\widetilde\phi}^\dagger\, \ell_L \right)
-\frac{1}{2}\left(  \overline{{N_R}^c} \, M \,{N_R} 
+\overline{N_R}\, M^* \,{N_R}^c \right)\,.
\label{LNR1}
\ee
${\cal L}_{N_R}$ contains kinetic energy and Majorana mass terms for the right-handed neutrinos as well as
Yukawa interactions between right-handed neutrino singlets,
left-handed lepton doublets $\ell_L$ and the Higgs boson scalar doublet.  
Since the right-handed neutrinos
$N_R$ are color and $SU(2)$ singlets with hypercharge
$Y=0$, the covariant derivative reduces to $D_\mu = \partial_\mu$ in the kinetic energy
terms.  In addition, lepton-number violating Majorana mass terms are allowed by the 
gauge symmetries.
The Majorana mass matrix $M$ is an $n'\times n'$
complex symmetric matrix 
with eigenvalues of ${\cal O({\cal M})}$.
Note that the charge conjugate of the chiral fermion field appearing in the Majorana mass term  is
defined by ${\psi_R}^c \equiv C \overline{\psi_R}^T$. 
The Yukawa interactions are written in terms of the Higgs boson doublet $\widetilde{\phi}$ which is related 
to the standard scalar doublet $\phi$ by $\widetilde{\phi} = i\tau_2 \phi^*$.
$Y_\nu$ is the $n \times n'$ matrix of neutrino Yukawa couplings.

The Majorana mass matrix $M$ has, in general, $n'$ complex eigenvalues
${M}_i=e^{i \theta_i}|M_i| \equiv \eta_i |M_i|$ which depend on the Majorana
phases $\theta_i$ of the heavy Majorana neutrinos.
We can work in the basis in which $M$ is real and diagonal. In this case, the Majorana neutrino mass eigenstates $N_i = N_i^c$
are given by
\be
N_i \equiv e^{i {\theta_i}/2} \, {N_R}_i + e^{-i {\theta_i}/2}\, {N_{R\,i}}^c
= \sqrt{\eta_i} \, {N_R}_i + \sqrt{\eta_i^*} \, {N_{R\,i}}^c\, ,
\ee
and the Lagrangian in Eq. (\ref{LNR1}) can be rewritten as
\bea
\label{lsinglet3}
{\cal L}_{N}&=& 
\frac{1}{2} \,\overline{N}_i \left(i\, \dv \,  -  M_i  \right) N_i \\ 
&-&  \frac{1}{2} \,\left[\,\overline{\ell_L} \, \widetilde\phi\, Y_\nu \sqrt{\eta^*} \,+
\overline{{\ell_L}^c} \,{\widetilde\phi}^*\, Y_\nu^* \,\sqrt{\eta} \, \right]_i\, N_i
- \frac{1}{2}\, \overline{N_i}\, \left[ \,\sqrt{\eta^*} \, Y_\nu^T \, 
{\widetilde\phi}^T \,{\ell_L}^c 
+ \,\sqrt{\eta} \, Y_\nu^\dagger\, {\widetilde\phi}^\dagger
\,\ell_L\right]_i~,\nn
\eea
where $\eta$ is the $n'\times n'$ diagonal matrix with elements
$\eta_i$. 
We adopt the convention that Latin indices denote 
mass eigenstates and Greek indices denote flavor eigenstates
throughout this paper.

An effective Lagrangian which is valid at energies less than ${\cal M}$ 
can be constructed by integrating out the heavy Majorana neutrino fields $N_i$.  
The effective Lagrangian has a power series expansion in 
$1/{\cal M}$ of the form 
\begin{equation}
\label{leff}
{\cal L}_{eff}= {\cal L}_{SM}+
     \frac{1}{{\cal M}}{\cal L}^{d=5} + \frac{1}{{\cal M}^2} {\cal L}^{d=6} +
     \cdots\  \equiv {\cal L}_{SM}+ {\cal \delta L}^{d=5} + {\cal \delta L}^{d=6} +
     \cdots\  ,
\end{equation}
where
${\cal L_{SM}}$ contains all
$SU(3) \times SU(2) \times U(1)$ invariant operators of dimension $d\leq 4$  and the gauge invariant operators
of dimension $d > 4$, constructed 
from the SM fields, account for the physics effects of the heavy Majorana neutrinos at energies   
$\le {\cal M}$.
The effective Lagrangian is defined through the effective action~\cite{Santa1},
\bea
\label{integraout}
e^{i S_{eff}}=
\exp\left\{ i \int d^4 x \,{\cal L}_{eff}(x) \right\} \equiv
\int {\cal D} N {\cal D} \overline{N} e^{i S} 
=e^{i S_{SM}}\int {\cal D} N{\cal D} \overline{N}  e^{i S_N} ~,
\eea
obtained by functional integration over the heavy Majorana neutrino fields.
The classical equations of motion for the $N$-field with solution $N_0$ are
obtained from
\be
\left.\frac{\delta S}{\delta N_i(x) } \right|_{{N_0}_i} =0~, \qquad
\left.\frac{\delta S}{\delta \overline{N}_i(x)}  \right|_{\overline{N_0}_i} =0~,
\ee
which yield
\bea
\label{sol}
\overline{N_0}_i\, ( -i {\dvl} -  M_i)
-\left( \overline{\ell_L}\,\widetilde\phi \,  Y_\nu \,\sqrt{\eta^*}+ 
 \overline{{\ell_L}^c}\, {\widetilde\phi}^* \, Y_\nu^* \,\sqrt{\eta}\,\right)_i & = &
0~, \\ \nn 
( i\dvr -M_i)\, {N_0}_i -
 \left( \sqrt{\eta} \, Y_\nu^\dagger \,{\widetilde\phi}^\dagger \,\ell_L + 
 \sqrt{\eta^*} \, Y_\nu^T \,{\widetilde\phi}^T \, {\ell_L}^c\right)_i & = & 0~.
\eea

The effective action is given by 
\be
\label{effact}
S_{eff}= S_{SM}+S_{N}[{N_0}]~,
\ee
where
\begin{equation}
\label{S0}
S_N[N_0] \approx -\frac {1} {2} 
 \int d^4x \left( \overline{\ell_L} \, 
\widetilde\phi \,Y_\nu\, \sqrt{\eta^*} 
+ \overline{{\ell_L}^c}\,\widetilde\phi^* \,  Y_\nu^* \sqrt{\eta} \, 
\,\right)_i \,
  \left(\frac{\delta_{ij}}{i\dvr -M_i}\right) \,\left(  \sqrt{\eta} \, Y_\nu^\dag\,
{\widetilde\phi}^\dag \,\ell_L 
+  \sqrt{\eta^*} \,Y_\nu^T \,{\widetilde\phi}^T \, {\ell_L}^c\,\right)_j\, .
\end{equation}

The high-energy Lagrangian ${\cal L}$ does not contain
loop corrections with only heavy neutrinos running around the loop, so
all contributions to the effective Lagrangian can be obtained by expanding 
the heavy neutrino propagator in a power series in $1/M$, 
\be
\label{expan}
\frac{1}{i\dvr -M}
= -\frac{1}{M}
- \frac{i\dvr}{M^2} +  \dots \, .
\ee
The substitution of Eq.~\rfn{expan} into \eq{S0} yields
the terms of dimension $\le 6$, which suffice for the purposes of this work.

\subsection{ d=5 operator}

Eq.~\rfn{S0} yields the $d=5$ operator of the effective Lagrangian for the seesaw model,
\be
\label{L52}
{\cal {\delta L}}^{d=5} = 
\frac{1}{2}\,  c^{d=5}_{\alpha \beta}\,
\left(\overline{{\ell_L}^c_\alpha}\,  {\widetilde\phi}^*\right)  \, 
 \left( {\widetilde\phi}^\dag \, {\ell_L}_\beta \right) 
+\frac{1}{2} \, \left({c^{d=5}_{\alpha \beta}}\right)^*\,
 \left( \overline{\ell_L}_\alpha \, \widetilde\phi \right) 
 \left( {\widetilde\phi}^T \, {\ell_L}^c_\beta \right) \ ,
\ee
where
\be
c^{d=5}_{\alpha \beta}=\left(Y_\nu^* 
\,\frac{\eta}{M} \,Y_\nu^\dagger \right)_{\alpha \beta}\,.
\ee

This expression can be rewritten in terms of $SU(2)$ singlet and  triplet components
of the leptons using a Fierz
identity. For the case under study the singlet term does not contribute 
because it is proportional to 
$\widetilde{\phi}^\dagger\,i \tau_2\,\widetilde{\phi}^*=0$.
Using the property 
$\overline{{\ell_L}^c} \widetilde{\phi}^* =\overline{{\ell_L}^c}\, i\tau_2\,
\phi \equiv  -\overline{\widetilde{\ell_L}} \,\phi$,
Eq.~(\ref{L52}) can be reexpressed as  
\bea
\label{dim5_def}
{\cal {\delta L}}^{d=5} = 
-\frac{1}{4} c^{d=5}_{\alpha \beta}\,\,
 \left( \overline{\widetilde{\ell_L}}_\alpha \, \vec{\tau} \, {{\ell_L}_\beta} \right) 
 \left( \widetilde{\phi}^\dagger \, \vec{\tau} \, {\phi}   \right)  + \mrm{h.c.}\, ,
\eea
which is the well-known $\left(\Delta L=2\right)$ $d=5$ operator~\cite{Weinberg} that generates Majorana masses 
for the light weak doublet
neutrinos $\nu_L$ when the Higgs doublet develops a
non-zero vacuum expectation value $v/\sqrt{2}\simeq 174 \ \gev$.  
The Majorana
mass matrix of the light neutrinos is given by
\be
\label{mass}
{m}_{\alpha\beta}= - \frac{v^2}{2}\left(Y_\nu^* 
\,\frac{\eta}{M} \,Y_\nu^\dagger \right)_{\alpha \beta}\, =
- \frac{v^2}{2} \left(c^{d=5}_{\alpha \beta}\right)\ .
\ee
Notice that the Majorana phases of the heavy singlet Majorana neutrinos are inherited by 
the Majorana mass matrix
of the light Majorana neutrinos.

\subsection{ d=6 operator}

\eq{S0} yields two $d=6$ operators
which can be shown to be identical, resulting in 
\be
\label{eq1}
{\cal {\delta L}}^{d=6} = i\,\left[ 
\overline{\ell_L} \,  \widetilde\phi\, Y_\nu \,   
\frac{\dvr}{M_i^2} \,\left(Y_\nu^\dagger\,{\widetilde\phi}^\dagger\,
\ell_L\right)\right]\, = c^{d=6}_{\alpha \beta} \left( \overline{\ell_L}_\alpha
\,  \widetilde\phi\right) i \dvr \left({\widetilde\phi}^\dagger\, {\ell_L}_\beta \right)\, ,
\ee
where
\be
c^{d=6}_{\alpha \beta} = \left( Y_\nu \frac{1}{M^2} 
Y_\nu^\dagger \right)_{\alpha \beta} \ .
\ee

The $d=6$ operator renormalizes the neutrino kinetic energy, which can be 
diagonalized through the redefinition
\be
\label{ren-wf}
\nu_\alpha^\prime =  
\left(\delta_{\alpha \beta}
+ \frac{v^2}{4} c^{d=6}_{\alpha
\beta} \right) \nu_\beta~\,,
\ee
where $\nu^\prime$ corresponds to the diagonal basis. 
Thus, one physical impact of the $d=6$ operator is to modify 
the couplings of neutrinos to gauge bosons.
It does not further modify the effects of the $d=5$ operator, since the effective Lagrangian
is restricted to ${\cal O}(1/{\cal M}^2)$ in this work.

It is instructive to rewrite the $d=6$ operator in a form which allows comparison with
the  tower of $d=6$ operators invariant under the SM found in the literature \cite{dim6}.
It is possible to promote the partial derivative to a covariant one,
\be
\partial^\mu ({\widetilde\phi}^\dagger \,\ell_L)= (\partial^\mu
{\widetilde\phi}^\dagger)\,\ell_L +
{\widetilde\phi}^\dagger\,\partial^\mu\ell_L \, =  
D^\mu ({\widetilde\phi}^\dagger \,\ell_L)= 
(D^\mu {\widetilde\phi}^\dagger)\,\ell_L +
{\widetilde\phi}^\dagger\,D^\mu\ell_L~,
\ee
since ${\widetilde\phi}^\dagger \,\ell_L$ is a gauge singlet.  In addition,
the equations of motion can be used to further simplify the operator. 
The equation of motion for the left-handed lepton doublet is given by
\bea
\label{eqmot}
 i\, \Dr\,\ell_L\,-\,Y_e\,\phi\,e_R\,+\,
\cdots = 0 
\eea
where $Y_e$ is the $n \times n$ matrix of the Yukawa couplings involving the right-handed charged leptons $e_R$. 
The ellipsis stands for terms suppressed in $1/{\cal M}$ 
which lead to higher dimension operators that are 
disgarded here. The term proportional to $Y_e$ does not contribute, since 
$\label{prop}
{\widetilde\phi}^\dagger \, \phi= -i \phi^T \,\tau_2\,\phi=0$. After a Fierz transformation, 
 \eq{eq1} can be rewritten as 
\be
\label{dim6_def}
{\cal {\delta L}}^{d=6}=  \frac{i}{2} \,c^{d=6}_{\alpha \beta}
\left\{\,(\overline{\ell_L}_\alpha \,\gamma_\mu\,{\ell_L}_\beta) \,
({\phi}^\dagger\,D_\mu \,\phi)\,
 -\,\frac{1}{2}(\overline{\ell_L}_\alpha
\,\vec\tau\,\gamma_\mu\,{\ell_L}_\beta)
\,\left(\phi^\dagger\,D_\mu\vec\tau\phi\,
+\,\phi^\dagger\,\vec\tau\,D_\mu\phi\right)\right\}~,
\ee
which contains the $SU(2)$ singlet and triplet operators largely
studied in the literature.
The singlet operator is well known to modify, when considered alone, 
the couplings of the $Z$ to neutrinos and 
charged leptons. The triplet operator by itself modifies the couplings of the $W$ and the $Z$. 
However, in the particular combination Eq.~\rfn{dim6_def}, corrections to the 
$Z$ couplings of charged leptons cancel and only the $W$ and $Z$ couplings which
involve neutrinos are modified. It can be easily verified that the physical consequences of 
Eq.~(\ref{dim6_def}) match those stemming from Eq.~(\ref{ren-wf}), as they must.

Inclusion of the $d=6$ operator in the charged
current implies that the leptonic mixing matrix of the effective theory is given by
\be
U^{eff}_{\alpha i} = \left( \delta_{\alpha\beta} 
- \frac{v^2}{4}\,c^{d=6}_{\alpha \beta}\right) \,U_{\beta i}\,, 
\ee
where $U$ is the usual MNS lepton mixing matrix. Thus, neutrino oscillations are affected by the presence of the
$d=6$ operator. We note that the sensitivity of neutrino oscillations 
to more phases than just the ``CKM''-like phase, although with effects 
suppressed by powers of 
$1/{\cal M}^2$,
has been pointed out already in Ref.~\cite{Endoh} in a general context. Phenomenological
bounds for the $d=6$ operator also can be found in the literature, as
this operator has been dealt with previously in the context of
theories with extra dimensions \cite{Gouvea}.
For the particular case of a very short baseline $L\simeq 0$, the oscillation
probability depends on the coefficient $c^{d=6}_{\alpha \beta}$
\cite{Gouvea}:
\be
P(\nu_\alpha\to\nu_\beta)= \left|\, \delta_{\alpha \beta} -\frac{v^2}{2}\,c^{d=6}_{\alpha\beta}
\,\right|^2\ .
\ee

>From the results of the short baseline experiments \cite{CHOOZ,SBL}, we obtain a 
bound
on the seesaw scale,
\be
  Y_\nu/M \lesssim 10^{-4} \ {\gev}^{-1} \,,
\ee
which is many orders of magnitude weaker than that obtained from the 
d=5 operator, although independent from it.

The analysis of the impact on observables of the imaginary part of the
$c^{d>4}$ coefficents is in progress \cite{more}.

\subsection{General Lagrangian and RG running}

While the $d=5$ operator of the effective Lagrangian
is the unique dimension-five operator
compatible with the 
gauge symmetries of the SM, there are many $d=6$ operators other than 
the one derived here.
This means that the effective Lagrangian
\be
{\cal L}_{eff}^{d\le6}=  {\cal L_{SM}} - \frac{1}{4}
\left[ c^{d=5}_{\alpha \beta}\,\,
 \left( \overline{\widetilde{\ell_L}}_\alpha \, \vec{\tau} \, {{\ell_L}_\beta} \right) 
 \left( \widetilde{\phi}^\dagger \, \vec{\tau} \, {\phi}   \right)  + \mrm{h.c.}\,\right]
+i \, c^{d=6}_{\alpha \beta}\,\,
\left[ \overline{\ell_L}_\alpha \,  \widetilde\phi\,  
\dvr\left({\widetilde\phi}^\dagger\,
{\ell_L}_\beta \right)\right]\,,
\label{effL}
\ee
is not the only possible form of the effective Lagrangian with $d\le6$
stemming from the general idea of the seesaw mechanism. More complicated 
scenarios for the heavy neutrino and scalar sectors, or further interactions, can add 
new operators with dimension $d=6$ to the 
Lagrangian \cite{more}.
Also, the RG evolution of the operator
couplings in Eq. (\ref{effL}) from the putative high-energy scale where 
they are produced down to the electroweak scale induces mixing
with other $d=6$ operators \cite{more} and changes the relationship of the $d=5$ coefficient
to the high-energy parameters. 
Nevertheless, unless very unnatural cancellations are present, 
our tree-level result should be 
a tell-tale signature of the seesaw mechanism.

\section{Parameter counting}

It is necessary and sufficient to consider our tree-level effective
Lagrangian Eq.~\rfn{effL} 
in order to take into account the leading low-energy 
effects related to leptogenesis. In this section, 
we show that the number of independent physical angles 
and $CP$-phases contained in our effective Lagrangian
with $d\le6$ 
equals that of the high-energy seesaw Lagrangian, when the number
of heavy and light neutrino generations is the same, $n'=n$.
The situation for $n'\ne n$ also is discussed.
We count how many physical parameters are contained in the effective Lagrangian by 
analyzing the symmetry structure of the theory, using the method developed 
in Ref.~\cite{Santa2}.

\subsection{Seesaw Lagrangian}

First consider the symmetry structure of the 
high-energy seesaw Lagrangian Eq. \rfn{Lagrangian} with $n$ light lepton
families and $n'$ right-handed Majorana neutrinos. The kinetic energy terms are 
invariant under the chiral transformations
\bea
\ell_L &\to& V_\ell \,\ell_L ~,\\
e_R &\to& V_e \,e_R  \nn~, \\
N_R &\to& V_N \, N_R  ~,
\eea
where $V_\ell$ and $V_e$ are $n\times n$ unitary matrices and $V_N$ is an 
$n'\times n'$ unitary matrix. The Yukawa sector and the
Majorana mass term explicitly
break the chiral invariance unless the coupling matrices
transform as
\bea
\label{transf2}
Y_e &\to& Y_e' \equiv V_\ell \,Y_e \, V_e^\dagger ~, \nn \\
Y_\nu &\to& Y_\nu' \equiv V_\ell \,Y_\nu \, V_N^\dagger  ~,\nn \\
M &\to& M' \equiv V_N^* \,M \, V_N^\dagger ~.
\eea
Eq.~(\ref{transf2}) defines an equivalence relation between theories with different
matrices 
\be
\label{relation}
(Y_e,Y_\nu,M) \leftrightarrow (Y_e' ,Y_\nu' ,M' ).
\ee

Counting how many physical parameters $N_{phys}$ are needed to describe the Yukawa
and Majorana mass
terms in the seesaw Lagrangian is tantamount to counting how many
equivalence classes there are with respect to the transformation \eq{relation}.
The result is given by 
\be
\label{nphys}
N_{phys}= N_{\rm order}-(N_G-N_{H}),
\ee
where $N_{\rm order}$ is the sum of the number of parameters contained in
the Yukawa and Majorana mass matrices, $N_{G}$ is the number of parameters contained in
the matrices of the 
chiral symmetry group $G=U(n)_\ell \times U(n)_e \times U(n')_N$  and $N_{H}$ is
the number of parameters contained in the matrices of the subgroup $H$ of 
the chiral symmetry group which remains unbroken by the Yukawa and Majorana mass matrices.  
In the seesaw model, the
chiral symmetry group is completely broken by the Yukawa and Majorana mass terms, so
there is no unbroken subgroup $H$.  The number of moduli and phases (real and imaginary
parameters) in the Yukawa and Majorana mass matrices and in the chiral symmetry matrices
$V_e$, $V_\ell$ and $V_N$ are tabulated in Table~1.  The number of physical parameters 
contained in the seesaw model is computed using Eq.~(\ref{nphys}), and appears at the bottom
of the table.  There are $(n + n^\prime + n n^\prime)$ physical moduli and $n(n^\prime -1)$ 
physical phases in the Yukawa and Majorana mass matrices of the seesaw model.  
Of the real parameters, $n$ are the charged
lepton masses, $n$ are the light Majorana neutrino masses and $n^\prime$ are the heavy
Majorana neutrino masses, whereas the remaining $(n n^\prime -n)$ real parameters 
are mixing angles.

It is useful to check this counting result for a few special cases.  
For $n = n^\prime =2$,
there are 8 moduli (6 masses and 2 mixing angles) and 2 $CP$-odd phases,
whereas for $n=n^\prime=3$, there are 15 moduli (9 masses and 6 mixing angles)
and 6 $CP$-odd phases. The case  $n=3$ and $n'=2$ results in 11
moduli (8 masses and 3 mixing angles) and 3 $CP$-odd phases.   

\begin{table}[ht]
\begin{center}
Table 1: Seesaw Model
\end{center}
\begin{center}
\begin{tabular}{|c|cc|}
\hline
Matrix &Moduli &Phases \\
\hline
$Y_e$ &$n\times n$ &$n\times n$\\
$Y_\nu$ &$n\times n'$ &$n\times n'$\\
$M$ &$\frac{n'(n'+1)}{2}$ &$\frac{n'(n'+1)}{2}$\\
\hline
$V_e$ &$\frac{n(n-1)}{2}$ &$\frac{n(n+1)}{2}$\\
$V_\ell$&$\frac{n(n-1)}{2}$&$\frac{n(n+1)}{2}$\\
$V_N$ &$\frac{n'(n'-1)}{2}$ &$\frac{n'(n'+1)}{2}$\\
\hline
$N_{phys}$ &$n+n'+ n n'$ &$n(n'-1)$ \\
\hline
\end{tabular}
\end{center}
\end{table}

\subsection{ Low-energy Effective Lagrangian}

Consider now the Lagrangian of the effective theory truncated at the $d=6$ operators
for $n$ light active families.
Consistency with the symmetries
of the high-energy Lagrangian implies that the coefficients of the $d=5$ and $d=6$
operators in the effective theory transform under the chiral symmetry as
\bea
\label{tranf2}
c^{d=5}&\to& \, 
V_\ell^* \, c^{d=5} \, V_\ell^\dagger\,, \\
c^{d=6} &\to& 
\,V_\ell\,c^{d=6}\,V_\ell^\dagger\,.
\eea

The chiral symmetry group of the effective theory is $G=U(n)_\ell \times U(n)_e$ 
since the heavy neutrinos have been integrated out of the theory. 
The $d=5$  term breaks the chiral
symmetry group $G$ completely, so there is no unbroken subgroup $H$.  
The real and 
imaginary parameters contained in the $d=5$ and $d=6$ operator
coefficient matrices are tabulated in Table 2, using the fact that
$c^{d=5}$ is an $n \times n$ complex symmetric matrix and 
$c^{d=6}$ is an $n \times n$ Hermitian matrix.  The number of
parameters contained in the $n \times n$ unitary matrices $V_e$ and $V_\ell$ 
of the chiral symmetry group also
are tabulated.  The number of physical parameters in the effective Lagrangian
up to operators of dimension six is given 
on the last line of Table 2.  The number of physical parameters in the low-energy effective
Lagrangian $(d\le 6)$ equals the number of physical parameters in the 
high-energy seesaw Lagrangian of Table~1 if $n^\prime = n$, demonstrating our assertion that all
of the physical parameters of the high energy theory appear in the low-energy effective theory
containing the $d=5$ and $d=6$ terms only. 
\begin{table}[ht]
\begin{center}
Table 2: Effective Theory ($d\le 6$)
\end{center}
\begin{center}
\begin{tabular}{|c|cc|}
\hline
Matrix &Moduli &Phases \\
\hline
$Y_e$ &$n\times n$ &$n\times n$ \\
$c^{d=5}$ &$\frac{n(n+1)}{2}$ &$\frac{n(n+1)}{2}$ \\
$ c^{d=6}$ &$\frac{n(n+1)}{2}$ &$\frac{n(n-1)}{2}$ \\
\hline
$V_e$ &$\frac{n(n-1)}{2}$ &$\frac{n(n+1)}{2}$\\
$V_\ell$ &$\frac{n(n-1)}{2}$ &$\frac{n(n+1)}{2}$ \\
\hline
$N_{phys}$ &$n(n+2)$ &$n(n-1)$ \\
\hline
\end{tabular}
\end{center}
\end{table}

Note that with an effective Lagrangian containing only the $d=5$ operator, 
information is lost: in this case,  
the number of physical moduli is $n(n+3)/2$ and the number of physical phases is 
$n(n-1)/2$, which does not equal the number of physical parameters of the high-energy seesaw
model for any value of $n^\prime$. 
For example, for $n=2$, the $d=5$ effective Lagrangian would contain only 
5 moduli 
(2 charged lepton masses, 2 neutrino masses and one mixing angle) and one phase, 
to be compared with the 8 moduli and 2 phases of the high-energy Lagrangian for
$n=n^\prime =2$. The addition of the $d=6$ operator allows to recover
the missing parameters.

When some extra symmetry or constraint is imposed in the high-energy Lagrangian(i.e., degenerate heavy neutrinos,  $n'<n$,  etc.), 
the low-energy Lagrangian still has the same form, which appears 
paradoxical since now it contains a larger number of independent parameters 
than the high-energy theory.  The resolution of the paradox is 
that hypothetical low-energy measurements then are correlated.

Consider for instance the special case of $m'$ degenerate neutrinos among the heavy $n'$ Majorana 
fields, $m'\le n'$. The number of physical parameters at
high energy decreases, as shown in Table 3\footnote{We thank A. Romanino for pointing out an error in our original  
 formulae in this Table.}.
\begin{table}[ht]
\begin{center}
Table 3: Degenerate Seesaw Model
\end{center}
\begin{center}
\begin{tabular}{|c|cc|}
\hline
&Moduli &Phases \\
\hline
$N_{phys}$ & $(n+ 1)\,(n'+ 1) - \frac{m'\,(m' +1)}{2}$ \qquad
&$n\,(n' -1)$ \\
\hline
\end{tabular}
\end{center}
\end{table}

For the simplest case $m'=n'=n$,
using the explicit form of the coefficients in the low-energy Lagrangian, Eqs.(\ref{dim5_def}) 
and (\ref{dim6_def}),
it is easy to show that if all the elements of the coefficient matrices, $c^{d=5}$ and 
$c^{d=6}$,
are determined experimentally, most of the information extracted from the
values of the $c^{d=6}$ elements is redundant with respect to the information 
resulting from 
the measurement of the $c^{d=5}$ elements. 
 For instance, for $n=n'=m'=2$, the seesaw
Lagrangian would contain 6 real parameters and 2 phases. After measuring 
all the coefficients of the $d=5$ operator, 5 moduli and 1 phase
are determined. Then, the supplementary knowledge of one modulus and one phase of the $c^{d=6}$
coefficient suffices to determine completely the value of the fundamental
parameters $M$ and $Y_\nu$ of the high-energy theory. 
Other models of particular interest will be presented elsewhere \cite{more}.

 Finally, for models in which $n'>n$, Tables 1 and 2 illustrate that 
the number of independent parameters of 
the seesaw model is larger than that of our effective $d\le 6$ 
Lagrangian, Eq.~(\ref{effL}). The leading low-energy effects are still given 
by the latter, as no new $d\le 6$ operators are generated. 
In order to fully take into account the remaining 
parameters of the high-energy theory, operators of $d>6$ have 
to be added to the effective 
Lagrangian \cite{more}.

\section{Conclusions}

We have established a generic relationship between the seesaw model,  
including its leptogenesis-related parameters, 
and exotic low-energy neutrino couplings.
The physical consequences of the low-energy dimension $6$ operator 
are suppressed by two inverse powers of the large seesaw scale, 
and consequently, there is 
little practical hope to observe them, 
unless the seesaw scale turns out to be surprisingly small. The present work allows, 
though, to quantify the difficulty of the task.

\section{Acknowledgments}
We thank A.~Donini and P.~Hern\'andez for illuminating discussions and reading the manuscript. 
M.B.G. and E.J. thank the Aspen Center for Physics for hospitality
during the initial stage of this work. We also thank A.~Romanino for pointing out an inconsistency in a formula 
in the original version of this paper. A.B acknowledges MECD for financial support by
FPU grant AP2001-0521. A.B and M.B.G were partially
supported by CICYT FPA2000-0980 project.
E.J. was supported in part by the Department
of Energy under grant DOE-FG03-97ER40546.

\end{document}